\documentclass[aps,onecolumn,amsfonts,eqsecnum,float,preprint]{revtex4} 
\usepackage{epsfig,amsmath}
\usepackage{bm}

\begin{document}

\title{Semiclassical approximation with zero velocity trajectories}

\author{ Yair Goldfarb, Ilan Degani and David J. Tannor} \affiliation{
Dept. of Chemical Physics, The Weizmann Institute of Science,
Rehovot, 76100 Israel}

\begin{abstract}

\noindent We present a new semiclassical method that yields an
approximation to the quantum mechanical wavefunction at a fixed,
predetermined position. In the approach, a hierarchy of ODEs are
solved along a trajectory with zero velocity. The new approximation
is local, both literally and from a quantum mechanical point of
view, in the sense that neighboring trajectories do not communicate
with each other. The approach is readily extended to imaginary time
propagation and is particularly useful for the calculation of
quantities where only \textit{local} information is required. We
present two applications: the calculation of tunneling probabilities
and the calculation of low energy eigenvalues. In both applications
we obtain excellent agrement with the exact quantum mechanics, with
a single trajectory propagation.

\end{abstract}

\maketitle


\section{Introduction}

\noindent
 The first semiclassical method, the WKB method\cite{wentzel}, was
published almost simultaneously with the publication of the
Schr\"odinger equation in 1926. Since then, semiclassical methods
have continued to attract great interest for two primary reasons.
First, semiclassical methods give insight into classical-quantum
correspondence. Second, for large systems they hold the promise of
significant computational advantages relative to full quantum
mechanical calculations. In particular, in recent years much
progress has been made in the chemical physics community in
developing \textit{time-dependent} semiclassical methods that are
accurate and efficient for multidimensional systems. By
semiclassical methods, one generally means the calculation of a
quantum mechanical wavefunction or propagator via the propagation of
classical (or classical-like) trajectories. Mathematically speaking,
semiclassical methods cast the time-dependent Schr\"odinger equation
(TDSE), which is a PDE, in terms of ODEs related to classical
equations of motion. From a physical point of view, the
semiclassical methods try to circumvent the non-locality of quantum
mechanics.

In this paper we present a new method that is to some extent a cross
between a numerical grid method and a semiclassical method. The
method is derived by inserting a trial form $e^{iS/\hbar}$ into the
TDSE and calculating the time-dependent complex phase at a
\textit{stationary} position. Since the method yields an approximate
solution of the TDSE at a fixed position it is a cousin of grid
methods.  However, we compute $S$ by solving a hierarchy of spatial
derivatives of $S$ along a stationary trajectory; information from
the neighboring trajectories is incorporated only through the
initial conditions. This property makes the approximation local from
the quantum mechanical point of view, and hence a cousin of
semiclassical methods. We refer to the new method as the
zero-velocity complex action method (ZEVCA) since it employs a
complex phase (action) and stationary trajectories.

The substitution of $e^{iS/\hbar}$ as an ansatz in the TDSE is the
same starting point as several methods that are based on the
hydrodynamic and Bohmian formulations of quantum
mechanics\cite{bohm,holland,bohmb,wyattb}. More specifically, ZEVCA
is related to the Derivative Propagation Method\cite{trahan}, the
Trajectory Stability Method\cite{liu2} and Bohmian Mechanics with
Complex Action\cite{goldfarb}. As in the case with ZEVCA, these
methods incorporate equations of motion for a hierarchy of spatial
derivatives of the phase (and amplitude) that are calculated along
trajectories. But unlike ZEVCA, in these other approaches the time
dependent trajectories \textit{propagate} in either real or complex
configuration space. The ZEVCA formulation employs fixed
trajectories that yield the time dependence of the wavefunction at a
\textit{fixed}, \textit{predetermined} position in configuration
space. In reference \cite{wyattb} section 7.2, Wyatt considers the
solution of the global hydrodynamic equations of quantum mechanics
on fixed grid points (Eulerian grid) but he dismisses its usefulness
as a numerical tool. The ZEVCA formulation shows how to obtain
useful output of a local propagation at a single grid point.

For a number of quantum mechanical calculations, such as thermal
rates or tunneling probabilities, knowledge of the wavefunction in
all of configuration space is unnecessary: these quantities can be
calculated by data at a single position or a small interval of
space. For such calculations, ZEVCA has a significant numerical
advantage since it produces local information at a predetermined
position. The first application of ZEVCA that we present in this
paper is the calculation of tunneling probabilities; ZEVCA is well
suited for this calculation since tunneling probabilities can be
calculated from a time integral of the probability current at a
fixed position. The second application is the calculation of the
lowest energy eigenvalue of a bound potential. For this application
we make a minor modification to the ZEVCA formulation to adapt it to
imaginary time propagation. In both applications we calculate the
results by propagating \emph{just a single trajectory}.

This paper is organized as follows. In section \ref{form} we
formulate ZEVCA for the real time Schr\"odinger propagator (section
\ref{schro}) and the imaginary time Schr\"odinger propagator
(section \ref{boltzmann}). Since the multi-dimensional
representation of ZEVCA is somewhat more complicated, we present
only the one-dimensional derivation. Section \ref{num_app} is
dedicated to the implementation of ZEVCA for the derivation of
tunneling probabilities (\ref{tunnel}) and the first energy
eigenvalue of a bound potential (\ref{lowene}). In section
\ref{summary} we present our summary and concluding remarks.


\section{Formulation}
\label{form}

\subsection{The ZEVCA real time Schr\"odinger propagator}
\label{schro}

\noindent
 We start by inserting the ansatz\cite{pauli,kurt,davidb}
\begin{equation}
\label{trial2} \psi(x,t)=\exp\left[\frac{i}{\hbar}S(x,t)\right],
\end{equation}
into the TDSE
\begin{equation}
i\hbar\psi_{t}=-\frac{\hbar^{2}}{2m}\psi_{xx}+V(x,t)\psi,
\end{equation}
where $S(x,t)$ is a complex function, $\hbar$ is Planck's constant
divided by $2\pi$, $m$ is the mass of the particle and $V(x,t)$ is
the potential energy function. The subscripts denote partial
derivatives. The result is a quantum Hamilton-Jacobi
equation\cite{pauli,kurt,davidb}
\begin{equation}
\label{st1}
 S_{t}+\frac{1}{2m}S^{2}_{x}+V=\frac{i\hbar}{2m}S_{xx},
\end{equation}
where we recognize the classical Hamilton-Jacobi equation on the
LHS. On the RHS is an additional non-classical term that can be
referred to as a "quantum potential". Our aim is to solve
eq.(\ref{st1}) using the method of characteristics. This method is
usually applied to the solution of first order PDEs. The underlying
idea of the method of characteristics is to transform a single PDE
(or a set of PDEs) to a set of ODEs that are solved along a
characteristic curve. The characteristic is defined by setting a
dependence between the independent variables of the PDE (or PDEs),
in our case $x$ and $t$. ZEVCA makes the simplest possible choice
for the characteristics:
\begin{equation}
\label{dxdt} \frac{dx}{dt}=0\ \Longrightarrow \ x(t)=x(0),
\end{equation}
i.e. the characteristics are trajectories that remain at a constant
position. The time dependence of the phase along the trajectory is
given by inserting $S(x,t)$ in the Lagrangian time derivative
defined as
\begin{equation}
\frac{d}{dt}\equiv\frac{\partial}{\partial
t}+\frac{dx}{dt}\frac{\partial}{\partial x}.
\end{equation}
The choice we made in eq.(\ref{dxdt}) equates the Lagrangian time
derivative and the partial time derivative
$\frac{d}{dt}=\frac{\partial}{\partial t}$. The result of the
substitution is
\begin{equation}
\label{dsdt}
\frac{dS}{dt}=S_{t}=\frac{i\hbar}{2m}S_{xx}-\frac{1}{2m}S^{2}_{x}-V,
\end{equation}
where we have used eq.(\ref{st1}). The integration of
eq.$(\ref{dsdt})$ requires $S_{x}[x(0),t]$ and $S_{xx}[x(0),t]$.
Note that the initial position $x(0)$ acts as a parameter. Defining
\begin{equation}
\label{def_sn} S_{n}[x(0),t]\equiv \left.\frac{\partial^{n}
S}{\partial x^{n}}\right|_{[x(0),t]},
\end{equation}
we can write a general equation of motion for $S_{n}[x(0),t] \
\forall n$ in the following manner. We take the $n^{th}$ spatial
partial derivative of eq.(\ref{st1})
\begin{equation}
\label{st2}
 (S_{t})_{n}+\frac{1}{2m}(S^{2}_{1})_{n}+V_{n}=\frac{i\hbar}{2m}S_{n+2},
\end{equation}
and insert the result in the time derivative of $S_{n}$
\begin{equation}
\label{dsndt}
\frac{dS_{n}}{dt}=(S_{t})_{n}=\frac{i\hbar}{2m}S_{n+2}-\frac{1}{2m}(S^{2}_{1})_{n}-V_{n},
\end{equation}
where we have assumed that the time and spatial derivatives are
interchangeable. Equation (\ref{dsndt}) reveals that the equation of
motion of $S_{n}$ depends on the result of two subsequent equations
(by the dependence on $S_{n+2}$) and on all prior equations by the
term $(S_{1}^{2})_{n}=\sum_{j=0}^{n}\binom{n}{j}S_{j+1}S_{n-j+1}$.
Hence, the general characteristic solution of eq.(\ref{st1}) yields
an infinite hierarchy of ODEs defined by eqs.(\ref{dsndt}) where
$n=0,1,...,\infty$. In other words, we have converted eq.(\ref{st1})
to an infinite set of local but coupled ODEs that generate the time
dependence of the complex phase and its spatial derivatives at
position $x(0)$. An approximation can now be obtained by truncating
the set at some $n=N$; this is done by setting $S_{N+1}=S_{N+2}=0$.
We summarize the equations of motion of the ZEVCA approximation
\begin{eqnarray}
\label{dxdt2}
\frac{dS_{n}}{dt}&=&\frac{i\hbar}{2m}S_{n+2}-\frac{1}{2m}(S^{2}_{1})_{n}-V_{n}[x(0)];
\ \ n=0,1,...,N,
\\ \nonumber
S_{N+1}&=&S_{N+2}=0,
\end{eqnarray}
where we emphasize that the partial derivatives of the potential are
taken at a fixed position $x(0)$. The initial conditions of
eqs.(\ref{dxdt2}) are given by
\begin{equation}
\label{initial}
S_{n}[x(0),0]=-i\hbar\left.\frac{\partial^{n}\ln[\psi(x,0)]}{\partial
x^{n}}\right|_{x(0)}.
\end{equation}
where we used the relation $S(x,0)=-i\hbar\ln[\psi(x,0)]$ from
ansatz (\ref{trial2}). After we solve set (\ref{dxdt2}), the
wavefunction at $x(0)$ is given by
\begin{equation}
\psi[x(0),t]=\exp\left[\frac{i}{\hbar}S_{0}[x(0),t]\right].
\end{equation}
%

\subsection{The ZEVCA imaginary time Schr\"odinger propagator}
\label{boltzmann}

\noindent
 The imaginary time Schr\"odinger propagator takes the form
$\exp\left[-\frac{\hat{H}\tau}{2}\right]$ where
\begin{equation}
\label{bh}
 \hat{H}=-\frac{\hbar^{2}}{2m}\frac{\partial^{2}}{\partial
x^{2}}+V(\hat{x}).
\end{equation}
Applying this operator to an initial function $\psi(x,0)$ yields a
"wavefunction" $\tilde{\psi}(x,\tau)$ that is a solution of a
Schr\"odinger-like equation
\begin{equation}
\label{the_wf}
 -2\tilde{\psi}_{\tau}(x,\tau)=-\frac{\hbar^{2}}{2m}\tilde{\psi}_{xx}(x,\tau)+V(x)\tilde{\psi}(x,\tau),
\end{equation}
where $\tau$ plays the role of time. The mapping of $\tau$ to pure
imaginary time $\tilde{t}=-\frac{i\hbar}{2}\tau$ (hence the
term---``imaginary time propagation") transforms the last equation
to the form of the TDSE
\begin{equation}
\label{sh_t}
i\hbar\tilde{\psi}_{\tilde{t}}(x,\tilde{t})=-\frac{\hbar^{2}}{2m}\tilde{\psi}_{xx}(x,\tilde{t})+V(x)\tilde{\psi}(x,\tilde{t}),
\end{equation}
where for simplicity we allow a slight misuse of notation
$\tilde{\psi}(x,\tau)\rightarrow\tilde{\psi}(x,\tilde{t})$. At this
stage we insert in eq.(\ref{sh_t}) an ansatz identical to
eq.(\ref{trial2})
\begin{equation}
\label{trial3}
\tilde{\psi}(x,\tilde{t})=\exp\left[\frac{i}{\hbar}\tilde{S}(x,\tilde{t})\right],
\end{equation}
and apply the ZEVCA formulation (section \ref{schro}). This yields
equations identical to eqs.(\ref{dxdt2}) and (\ref{initial}) with
the only difference being $t\rightarrow\tilde{t}$.


\section{Model applications}
\label{num_app}

\subsection{Tunneling probabilities}
\label{tunnel}

\noindent In this section we calculate tunneling probabilities form
an Eckart barrier using a single trajectory propagation. The
potential is given by
\begin{equation}
\label{bv}
V(x)=\frac{D}{[\cosh(\beta x)]^{2}},
\end{equation}
where $D$ is the barrier height and $1/\beta$ gives an estimate of
the barrier width. The initial wavefunction is a Gaussian wavepacket
\begin{equation}
\label{gaussian}
\psi(x,0)=\exp\left[-\alpha_{0}
 (x-x_{c})^{2}+\frac{i}{\hbar}p_{c}(x-x_{c})+\frac{i}{\hbar}\gamma_{0}\right],
\end{equation}
where $1/\sqrt{\alpha_{0}}$ relates to the Gaussian width and
$\gamma_{0}=-\frac{i \hbar}{4}\ln(\frac{2\alpha_{0}}{\pi})$ takes
care of the normalization. $x_{c}$ and $p_{c}$ are the average
position and momentum respectively. The initial conditions of
eqs.(\ref{dxdt2}) are obtained by inserting eq.(\ref{gaussian}) in
eqs.(\ref{initial})
\begin{eqnarray}
\label{moment0}
S_{0}(x,0)&=&i\alpha_{0}\hbar(x-x_{c})^{2}+p_{c}(x-x_{c})+\gamma_{0},
 \\ \nonumber
 S_{1}(x,0)&=&2i\alpha_{0}\hbar(x-x_{c})+p_{c}, \\ \nonumber
 S_{2}(x,0)&=&2i\alpha_{0}\hbar, \\ \nonumber
 S_{n}(x,0)&=&0, \ n\geq 3.
\end{eqnarray}

The final tunneling probability $T$ for an initial wavefunction
centered at $x_{c}\ll0$ and having $p_{c}\geq0$ is given by
\begin{equation}
\label{bt} T=\lim_{t\rightarrow\infty}T(t),
\end{equation}
where
\begin{equation}
\label{bbt} T(t)=\int_{0}^{\infty}|\psi(x',t)|^{2}dx',
\end{equation}
is the time-dependent tunneling probability. The integration
initiates at $x=0$ since this is the position of the maximum of the
barrier. $T(t)$ can be expressed by a time integration of the
probability current at $x=0$. We show this by first writing the
quantum mechanical continuity equation
\begin{equation}
\label{cont} \rho_{t}(x,t)=-J_{x}(x,t),
\end{equation}
where $\rho\equiv|\psi|^2$ is the probability amplitude and the
probability current $J$ is given by
\begin{equation}
\label{bj} J=\frac{\hbar}{m}\Im(\psi^{\dag}\psi_{x}).
\end{equation}
By inserting $\rho(x,t)$ is eq.(\ref{bbt}) we can write
\begin{eqnarray}
\label{tt} T(t)&=&\int_{0}^{\infty}\rho(x',t)dx' \\ \nonumber
 &=&\int_{0}^{\infty}dx'\int_{0}^{t}\rho_{t}(x',t')dt' \\ \nonumber
 &=&-\int_{0}^{\infty}dx'\int_{0}^{t}J_{x}(x',t')dt' \\ \nonumber
 &=&-\int_{0}^{t}dt'\int_{0}^{\infty}J_{x}(x',t')dx'=\int_{0}^{t}J(0,t')dt'. \\ \nonumber
\end{eqnarray}
where in the third stage we used eq.(\ref{cont}) and in the forth
stage we changed the order of the integration and performed the
spatial integration. Inserting ansatz (\ref{trial2}) in
eq.(\ref{bj}) reveals that
\begin{equation}
\label{j2}
J=\frac{|\psi|^{2}}{m}\Re(S_{1})=\frac{1}{m}\exp\left[\frac{-2\Im(S_{0})}{\hbar}\right]\Re(S_{1}),
\end{equation}
where we have used the notation defined by eq.(\ref{def_sn}).
Inserting eq.(\ref{j2}) in the final result of eq.(\ref{tt}) yields
\begin{equation}
\label{btt}
 T(t)=\frac{1}{m}\int_{0}^{t}
\exp\left\{\frac{-2\Im[S_{0}(0,t')]}{\hbar}\right\}\Re[S_{1}(0,t')]dt'.
\end{equation}
Note that $S_{1}$ is readily obtained by the propagation of set
(\ref{dxdt2}). To calculate $T(t)$ we need to set $N$ and solve
eqs.(\ref{dxdt2}) with initial conditions (\ref{initial}) at
$x(0)=0$. Inserting $S_{0}(0,t)$ and $S_{1}(0,t)$ in eq.(\ref{btt})
completes the derivation of the tunneling probability.

\begin{figure}[h]
\begin{center}
\epsfxsize=9.5 cm \epsfbox{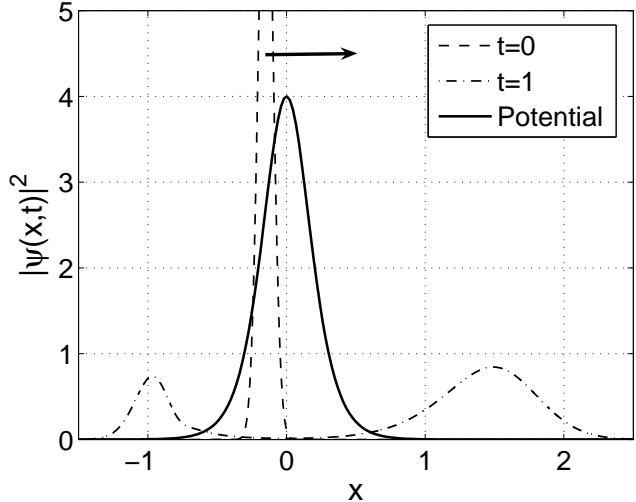} 
\end{center}
\caption{\label{wf} Plot of an initial Gaussian wavefunction
propagating in an Eckart barrier. At $t=1$ where the wavefunction is
clearly divided into a reflected part and a transmitted part. The
depiction of the barrier's width is in proportion to that of the
wavefunction, whereas the height of the barrier has no physical
meaning. The arrow indicates the direction of the average average
momentum. The parameters of the system are given in the text.}
\end{figure}

We turn to the numerical results. The parameters we insert in
eqs.(\ref{bv}) and (\ref{gaussian}) are $D=40$, $\beta=4.3228$,
$\alpha_{0}=30\pi$, $x_{c}=-1.5$, and $p_{c}=\sqrt{2mE}$ where
$E=20$ and $m=30$. All quantities here and henceforth are given in
atomic units, hence $\hbar=1$. In fig.\ref{wf} we illustrate the
potential and the wavefunctions. Note that the initial wavefunction
is located close to the barrier maximum, for reasons that we discuss
below.

\begin{figure}[h]
\begin{center}
\epsfxsize=9.5 cm \epsfbox{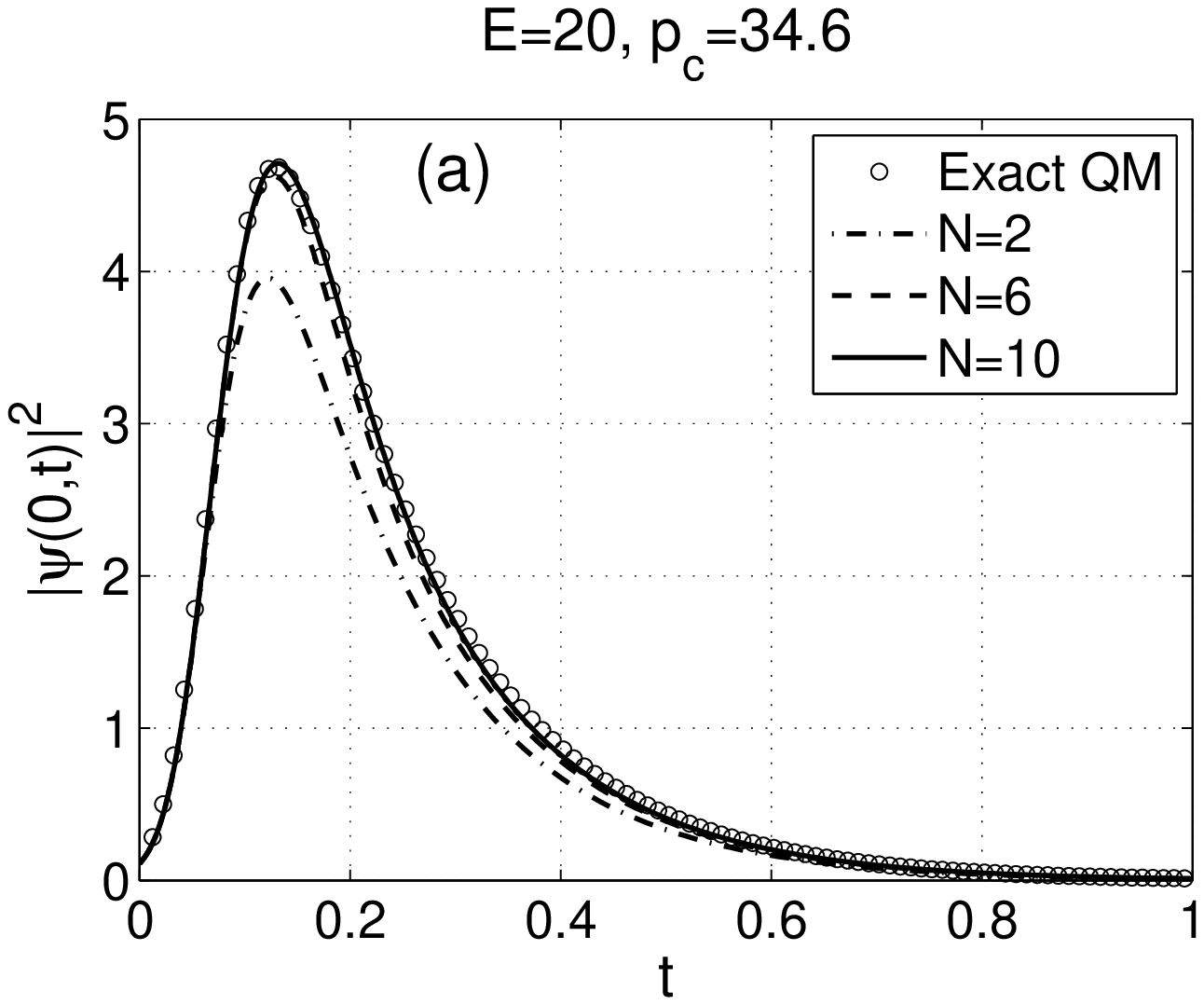} 
\epsfxsize=9.5 cm \epsfbox{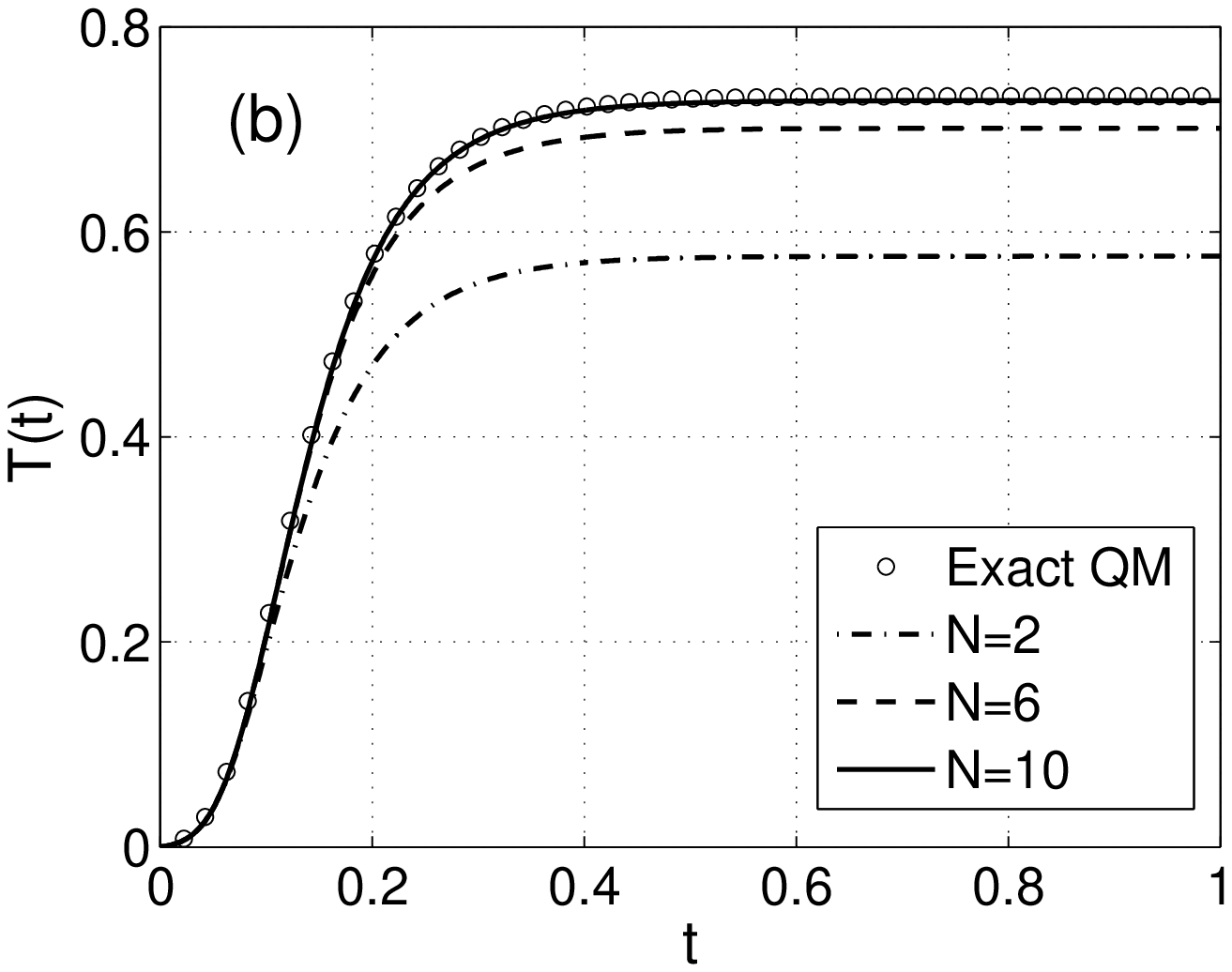} 
\end{center}
\caption{\label{e20} ZEVCA numerical results for $|\psi(0,t)|^{2}$
((a)) and the transmitted probability $T(t)$ ((b)) for a series of
values of the truncation order $N$. The system corresponds to
fig.\ref{wf} where the exact parameters are given in the text. The
relative error between the exact tunneling probability and the
asymptotic value of $T(t)$ (see eq.(\ref{bt})) for $N=2$, $N=6$, and
$N=10$ is roughly $20\%$, $4\%$ and $0.5\%$ respectively.}
\end{figure}

\begin{figure}[h]
\begin{center}
\epsfxsize=9.5 cm \epsfbox{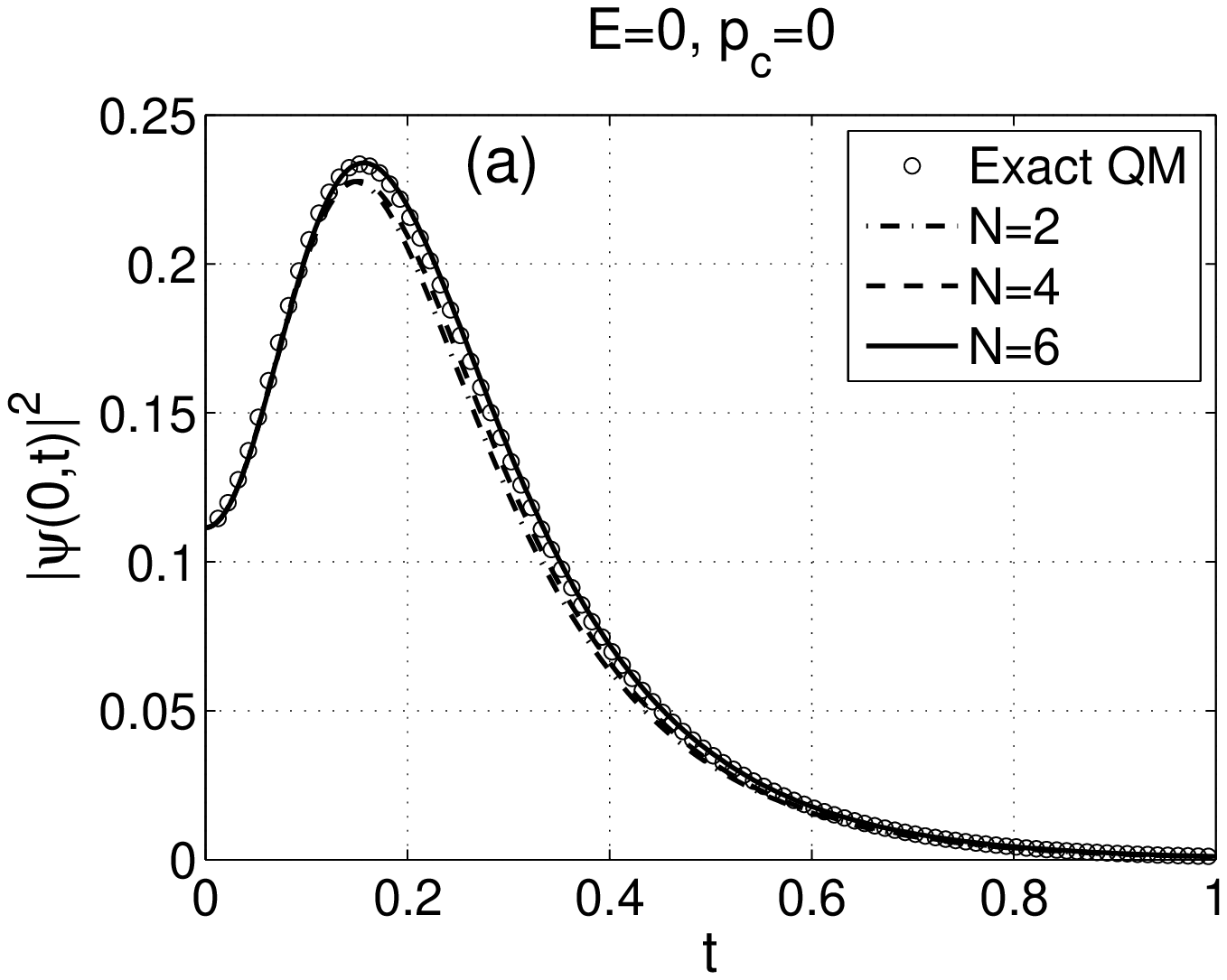} 
\epsfxsize=9.5 cm \epsfbox{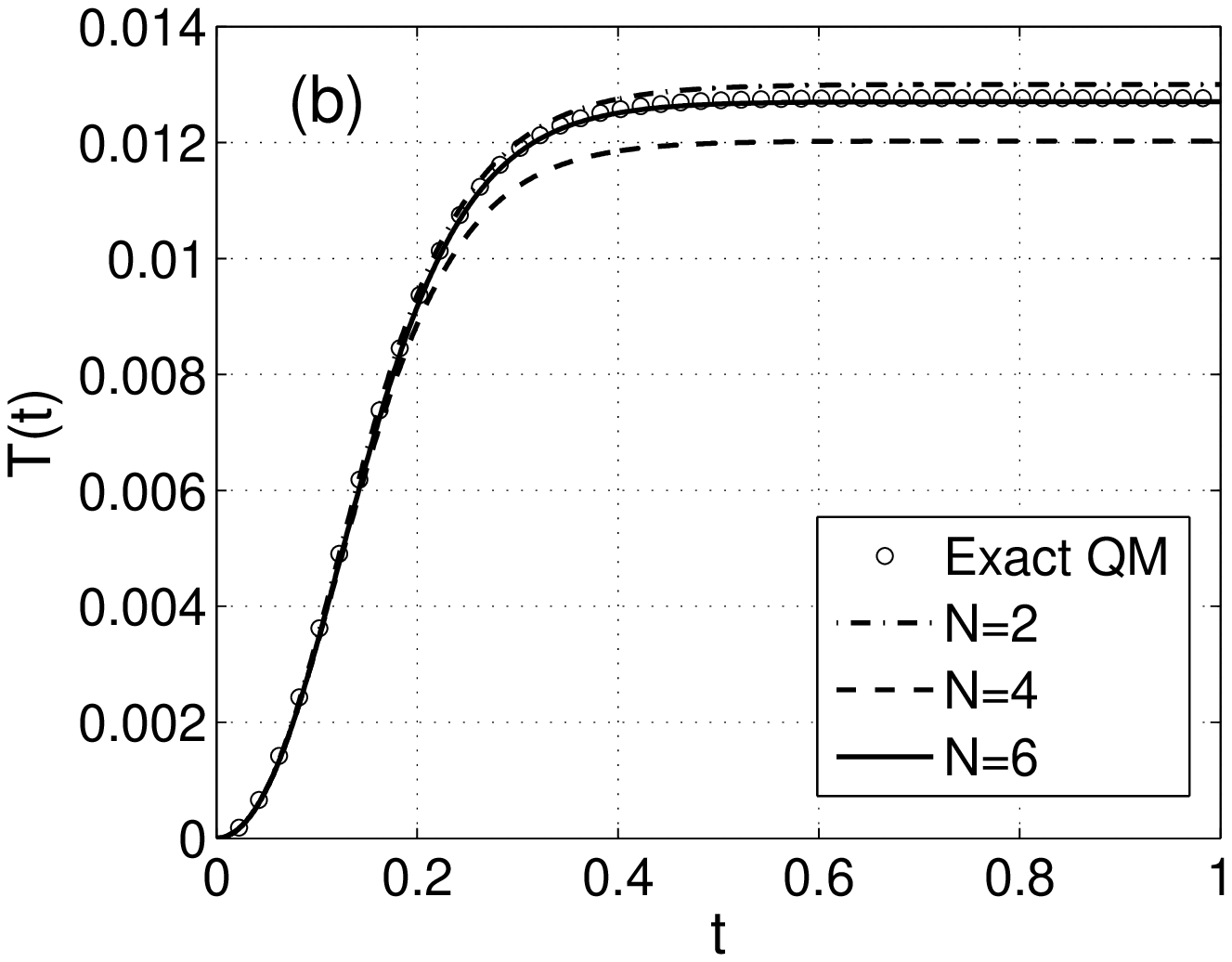} 
\end{center}
\caption{\label{e0} Same as fig.\ref{e20} expect that $p_{c}=0$. The
relative errors between the the exact tunneling probability and the
asymptotic value of $T(t)$ (see eq.(\ref{bt})) for $N=2$, $N=4$, and
$N=6$ are roughly $1.5\%$, $6\%$ and $0.8\%$ respectively.}
\end{figure}

In figures \ref{e20}(a) and \ref{e20}(b) we depict the
approximations to $|\psi(0,t)|^{2}$ and $T(t)$ for a series of
values of the truncation order $N$. Note that $|\psi(0,t)|^{2}$ is
equal to the exponential term in eq.(\ref{btt}). The relative error
between the exact tunneling probability and the asymptotic value of
$T(t)$ (see eq.(\ref{bt})) for $N=2$, $N=6$, and $N=10$ is roughly
$20\%$, $4\%$ and $0.5\%$ respectively. Clearly, the numerical
results converge quickly to the exact quantum mechanical result. In
figures \ref{e0}(a) and \ref{e0}(b) we plot $|\psi(0,t)|^{2}$ and
$T(t)$ for a set of $N$'s where we take $p_{c}=0$. Although the
relative error in the wavefunction $\psi(0,t)$ still converges
uniformly, the relative error in $T(t)$ does not: the errors are
$1.5\%$, $6\%$ and $0.8\%$ for $N=2$, $N=4$, and $N=6$ respectively.

The ZEVCA formulation of tunneling probabilities has a significant
restriction on its use. Since the ZEVCA trajectories remain fixed,
the formulation is very sensitive to the initial conditions inserted
in the equations of motion. The choice of $x(0)$ must satisfy two
properties: (1) The derivatives of the potential ($V_{n}[x(0)]$,
$n=1,...,N$) must have at least one term that is significantly
different from zero. (2) The initial wavefunction at $x(0)$
($\psi[x(0),0]$) needs to be significantly different from zero.
These restrictions ensure that the initial conditions "capture" the
wavefunction and the potential's surroundings at $x(0)$. These
restrictions have prevented us from choosing an $x_{c}$ in the
negative asymptotic region of the Eckart barrier since such a choice
would have produced $\psi(0,0)\rightarrow0$.


\subsection{Energy eigenvalues}
\label{lowene}

\noindent
 The ZEVCA imaginary time propagator allows for a simple calculation of
the first energy eigenvalue of a bound potential. As in the pervious
application, the calculation requires just a single trajectory
propagation. We start with a short derivation that demonstrates how
the first eigenvalue may be calculated using imaginary time
propagation. An arbitrary bound potential defines a set of
eigenfunctions $\phi_{j}(x),\ j=1,2...,\infty$ that satisfy
$\hat{H}\phi_{j}(x)=E_{j}\phi_{j}$, where $E_{j}$ are the energy
eigenvalues and $\hat{H}$ is given in eq.(\ref{bh}). The
eigenfunctions can be used as a basis set for the expansion of an
arbitrary wavefunction $\psi(x)$
\begin{equation}
\label{sum}
 \psi(x)=\sum_{j=1}^{\infty}a_{j}\phi_{j}(x),
\end{equation}
where $\sum_{j=0}^{\infty}|a_{j}|^{2}=1$. If we define a time scale
$\tau_{1}=\frac{2\pi\hbar}{E_{1}}$ then the operation of the
imaginary time propagator for $\tau \gg \tau_{1}$ on an initial
wavepacket $\psi(x,0)$ yields
\begin{eqnarray}
\label{derv}
\lim_{\tau\gg\tau_{1}}\exp\left[-\frac{\hat{H}\tau}{\hbar}\right]\psi(x,0)&=&
\lim_{\tau\gg\tau_{1}}\exp\left[-\frac{\hat{H}\tau}{\hbar}\right]\sum_{j=1}^{\infty}a_{j}\phi_{j}(x)
\\ \nonumber
&=&\lim_{\tau\gg\tau_{1}}\sum_{j=1}^{\infty}a_{j}\phi_{j}(x)\exp\left[-\frac{E_{j}\tau}{\hbar}\right]
\\ \nonumber
&=&a_{1}\phi_{1}(x)\exp\left[-\frac{E_{1}\tau}{\hbar}\right].
\end{eqnarray}
In the first stage we inserted eq.(\ref{sum}), in the second stage
we applied the imaginary time propagator on the eigenfunctions and
in the last stage we applied the limit. Inserting the relation
$\tau=\frac{2i}{\hbar}\tilde{t}$ in the final result of
eq.(\ref{derv}) and equating it with eq.(\ref{trial3}) yields
\begin{equation}
a_{1}\phi_{1}(x)\exp\left[-\frac{2iE_{1}\tilde{t}}{\hbar^{2}}\right]=
\exp\left[\frac{i}{\hbar}\tilde{S}(x,\tilde{t})\right],
\end{equation}
hence,
\begin{equation}
\tilde{S}(x,\tilde{t})=-\frac{2E_{1}\tilde{t}}{\hbar}-i\hbar\ln[a_{1}\phi_{1}(x)].
\end{equation}
Taking the $\tilde{t}$ partial derivative of the last equation
yields
\begin{equation}
E_{1}=-\frac{\hbar}{2}\tilde{S}_{\tilde{t}}(x,\tilde{t}),
\end{equation}
where we recall that this relation holds for
$\tau\gg\tau_{1}\Longrightarrow|\tilde{t}|\gg\hbar\tau_{1}$. As we
mentioned in section \ref{boltzmann}, eqs.(\ref{dxdt2}) hold for the
imaginary time propagator with the substitution of
$t\rightarrow\tilde{t}$. Hence,
\begin{equation}
\label{e1}
E_{1}=-\frac{\hbar}{2}\tilde{S}_{\tilde{t}}=-\frac{\hbar}{2}\frac{d\tilde{S}}{d\tilde{t}}=
-\frac{\hbar}{2}\left[\frac{i\hbar}{2m}\tilde{S}_{2}-\frac{1}{2m}(\tilde{S}_{1})^{2}-V\right].
\end{equation}
Since this equation holds for every choice of $x(0)$ we need to
propagate only a single trajectory for a sufficiently long
(imaginary) time $\tilde{t}$ to calculate $E_{1}$.

Liu and Makri have used a Bohmian related formulation, the
trajectory stability method (TSM), to calculate energy eigenvalues
\cite{liu1}. TSM \cite{liu2} emerges from conventional Bohmian
mechanics by constructing a hierarchy of equations of motion for
spatial derivatives of the phase and the amplitude of the
wavefunction. In reference \cite{liu1}, Liu and Makri use TSM for
imaginary time propagation at constant-position characteristics, as
we do here. But the modification of TSM for imaginary time
propagation is non-unique; moreover, producing constant-position
characteristics is quite an elaborate procedure that must be
repeated at every time step. In contrast, the ZEVCA transformation
from the Schr\"odinger real time propagator to the imaginary time
propagator is unique and the fixed characteristics are obtained for
all time simply by \textit{choosing} $\frac{dx}{dt}=0$. As we
demonstrate, the energy eigenvalues obtained using ZEVCA are of the
same accuracy as using TSM while the formulation is decidedly
simpler.

\begin{figure}[h]
\begin{center}
\epsfxsize=9.5 cm \epsfbox{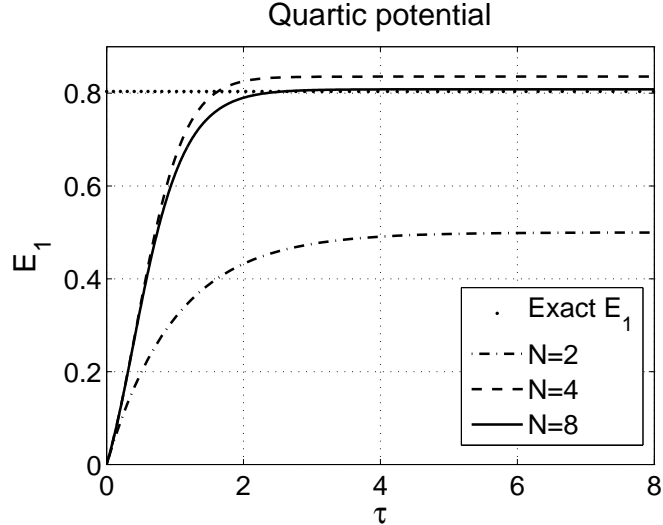} 
\end{center}
\caption{\label{x2x4} A comparison between the exact lowest energy
eigenvalue of a quartic potential and the results obtained using a
the ZEVCA imaginary time propagator in eq.(\ref{e1}) with series of
values of the truncation parameter $N$. The potential function is
$V(x)=\frac{1}{2}x^{2}+x^{4}$, where the numerical parameters appear
in the text. The relative error between the exact energy eigenvalue
($E_{1}=0.8038$) and the ZEVCA approximations for $N=2$, $N=4$ and
$N=8$ is roughly $38\%$, $4\%$ and $0.6\%$ respectively.}
\end{figure}

For the sake of comparison, we consider two of the potentials that
were studied in reference \cite{liu1}. The first is a quartic
potential $V(x)=\frac{1}{2}x^{2}+x^{4}$. The parameters of the
Gaussian initial wavepacket (eq.(\ref{gaussian})) are
$\alpha_{0}=0.5$, $x_{c}=1$ and $p_{0}=0$ with $m=1$. The fixed
trajectory that we propagate is at $x(0)=0.$ The results are very
robust with respect to the choice of the initial parameters. In
fig.\ref{x2x4} we depict $E_{1}$ (eq.(\ref{e1})) as function of
$\tau$ ($\tau=\frac{2i}{\hbar}\tilde{t}$) for a series of values of
the truncation order $N$. The relative error between the exact
energy eigenvalue ($E_{1}=0.8038$) and the ZEVCA approximations are
roughly $38\%$, $4\%$ and $0.6\%$ for $N=2$, $N=4$ and $N=8$
respectively. For $N=16$ the relative error reaches $0.1\%$. We see
a clear convergence to the exact quantum mechanical result as a
function of $N$.

\begin{figure}[h]
\begin{center}
\epsfxsize=9.5 cm \epsfbox{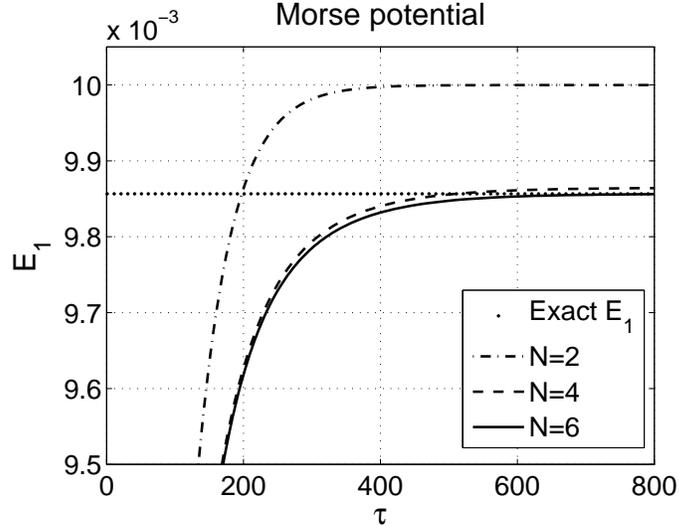} 
\end{center}
\caption{\label{morse} A comparison between the exact lowest energy
eigenvalue of a Morse potential and the results obtained by using
the ZEVCA imaginary time propagator in eq.(\ref{e1}) with series of
values of the truncation parameter $N$. The Morse potential
parameters correspond to the vibration of an H$_{2}$ molecule (the
numerical parameters appear in the text). The plot is a zoom of a
complete graph that initiates at $\tau=0$,
$E_{1}\simeq4.5\cdot10^{-3}$. The relative error between the exact
energy eigenvalue ($E_{1}=0.0098565$) and the ZEVCA approximations
is roughly $1.4\%$, $0.08\%$ and $2\cdot10^{-4}\%$ for $N=2$, $N=4$
and $N=6$ respectively.}
\end{figure}

The second example is a Morse potential $V(x)=D[1-\exp(-\alpha
x)]^{2}$. The parameters of the potential and the mass are
$D=0.1745$, $\alpha=1.026$ and $m=1837/2$. These parameters
correspond to the vibration of an H$_{2}$ molecule. The parameters
of the initial Gaussian wavepacket (eq.(\ref{gaussian})) are
$\alpha_{0}=4.5924$, $x_{c}=0.1$ and $p_{0}=0$. The fixed trajectory
is positioned at $x(0)=0$. As in the previous example, the results
are robust with respect to the choice of the initial Gaussian
parameters and the position of the fixed trajectory, $x(0)$. In
fig.\ref{morse} we depict $E_{1}$ as function of $\tau$ for a series
of values of the truncation order $N$. The relative error between
the exact energy eigenvalue ($E_{1}=0.0098565$) and the ZEVCA
approximations are roughly $1.4\%$, $0.08\%$ and $2\cdot10^{-4}\%$
for $N=2$, $N=4$ and $N=6$ respectively. In this example we see a
faster convergence to the exact result as a function of $N$ than in
the quartic case. The reason is that the parameters of the Morse
potential correspond to a small perturbation from a harmonic
oscillator potential. For the harmonic oscillator it is possible to
show that the ZEVCA approximation for $N=2$ yields the exact quantum
result. For anharmonic oscillators, the truncation at $N=2$
incorporates only $V_{j}[x(0),0], \ j=0,1,2$ in eqs.(\ref{dxdt2}),
which is equivalent to making a harmonic approximation to the
potential (note that the results obtained for $N=2$ for both the
quartic potential and the Morse potential correspond to the energy
eigenvalue of the harmonic approximation to the two potentials
respectively). Since the parameters of the Morse potential
correspond to a smaller perturbation from a harmonic oscillator
potential than the parameters for the quartic potential, the
convergence with $N$ is faster.
%


\section{Summary}
\label{summary}

\noindent We have presented ZEVCA, a new approximation for quantum
dynamics calculations that is a cross between a grid method and a
semiclassical method. The formulation was applied to the calculation
of tunneling probabilities and low energy eigenvalues, with
surprisingly good accuracy. The ZEVCA formulation has several
advantages: (1) The derivation of the formulation is straightforward
and the equations of motion are readily solvable by standard
numerical software. (2) The ZEVCA approximation yields the solution
of the TDSE at a fixed and predetermined position. This allows for
easy application to quantum quantities that require only local
information. (3) The ZEVCA formulation requires the calculation of
the potential and its derivatives only at $x(0)$ (see
eqs.(\ref{dxdt2})). This is in contrast to most semiclassical
methods that propagate trajectories in configuration space and
require the calculation of the potential (or its derivatives) at
each time step. (4) No root search is needed as in many other
semiclassical methods. (5) The extension to imaginary time
propagation is easily attainable. (6) By taking
$N\rightarrow\infty$, ZEVCA formally gives the exact quantum result.
Although we have not conducted a rigorous comparison of timings, for
the applications presented here ZEVCA was found to be several orders
of magnitude faster than the exact quantum calculation using a Split
Operator method.

Still, the numerical implementations reveals a number of
limitations. First, the convergence to the exact result as a
function of $N$ seems to be asymptotic, in the sense that there is
an optimal choice of $N$. Second, the method has difficulty at nodal
positions. Surprisingly, this is not a result of the original ansatz
(eq.(\ref{trial2})) but is a limitation imposed by the condition of
fixed trajectories. Since the trajectories are fixed, clearly they
cannot cross. We have demonstrated elsewhere that making the ansatz
(\ref{trial2}) but allowing for contributions from crossing
trajectories can produce interference effects such as nodes and
oscillations\cite{goldfarb2}. Third, in section \ref{tunnel} we have
presented limitations on the relation between the position of the
initial wavefunction, the derivatives of the potential and the
choice of $x(0)$. These limitations originate from the need to
incorporate in the initial conditions of ZEVCA "sufficient" data for
a successful propagation.

An alternative to the procedure we have described is to discard
ansatz (\ref{trial2}) and construct a hierarchy of ODEs for the
wavefunction $\psi_{n}[x(0),t]$ itself, instead of for the complex
phase, $S_{n}[x(0),t]$. We have explored this direction but found
that it produces very poor results. In the case of an initial
Gaussian wavepacket it is readily verified that the truncation for
$N\geq2$ does not entail any approximation to the complex phase
derivatives $S_{n}[x(0),0]$ (see eqs.(\ref{moment0})). This is not
the case for the derivatives of the initial wavefunction
$\psi_{n}[x(0),0]$ itself, which go to infinity as a function of $N$
for every choice of $x(0)$. This observation provides additional
justification for making the replacement $\psi \rightarrow
e^{iS/\hbar}$ in the first place.

We are currently working on several extensions of the ZEVCA
formulation. Specifically, we are exploiting the local properties of
ZEVCA for the calculation of thermal rates\cite{goldfarb3}. Similar
to the implementations of ZEVCA we have presented in this paper, the
calculation of thermal rates requires only local information. This
allows us to obtain thermal rates for one-dimensional systems using
only two trajectory propagations. We aim at exploring both ZEVCA and
its close relative, Bohmian Mechanics with Complex Action, in
multi-dimensional quantum systems.

This work was supported by the Israel Science Foundation (576/04).

%

\end{document}